\documentclass[acmsmall,nonacm]{acmart}

\setcopyright{none}
\renewcommand\footnotetextcopyrightpermission[1]{}
\usepackage{graphicx}
\usepackage{booktabs}      
\usepackage{multirow}
\usepackage{threeparttable}
\usepackage{array}
\usepackage{amsmath}
\usepackage{algorithm}
\usepackage[noend]{algpseudocode}
\usepackage{xspace}
\usepackage{listings}
\usepackage{xcolor}        
\usepackage{enumitem}
\usepackage{pifont}
\usepackage{bbding}
\usepackage{xfrac}
\usepackage{diagbox}
\usepackage{float}
\usepackage{ragged2e}
\usepackage{balance}       

\algrenewcommand\textproc{\texttt}

\usepackage{pifont}
\usepackage{times}
\usepackage{algorithm}
\usepackage{algorithmicx}
\usepackage{algpseudocode}
\usepackage{latexsym}
\usepackage{longtable}
\usepackage{threeparttablex}
\usepackage{amsmath}
\usepackage{multirow}

\newcommand{\toolname}[0]{\textsc{UIFormer}}

\definecolor{DarkGreen}{RGB}{1,100,32}
\definecolor{DarkRed}{RGB}{158,19,22}
\usepackage{pifont}

\usepackage{tikz}
\definecolor{customblue}{RGB}{192, 208, 235}
\definecolor{customdarkblue}{RGB}{68, 114, 196}

\AtBeginDocument{%
  \providecommand\BibTeX{{%
    \normalfont B\kern-0.5em{\scshape i\kern-0.25em b}\kern-0.8em\TeX}}}

\renewcommand\footnotetextcopyrightpermission[1]{}

\begin{document}

\title[\toolname{}]{From User Interface to Agent Interface: Efficiency Optimization of UI Representations for LLM Agents}

\author{Dezhi Ran}
\orcid{0000-0002-7916-255X}
\affiliation{%
  \institution{Key Lab of HCST (PKU), MOE; SCS, Peking University}
  \city{Beijing}
  \country{China}
}
\email{dezhiran@pku.edu.cn}

\author{Zhi Gong}
\orcid{}
\affiliation{%
  \institution{Tencent Inc.}
  \city{Shenzhen}
  \country{China}
}
\email{davidgong@tencent.com}

\author{Yuzhe Guo}
\orcid{0009-0000-7533-1478}
\affiliation{%
  \institution{Key Lab of HCST (PKU), MOE; SCS, Peking University}
  \city{Beijing}
  \country{China}
}
\email{yuzhe.guo@bjtu.edu.cn}

\author{Mengzhou Wu}
\orcid{0009-0004-0300-0994}
\affiliation{%
  \institution{Key Lab of HCST (PKU), MOE; SCS, Peking University}
  \city{Beijing}
  \country{China}
}
\email{wmz@stu.pku.edu.cn}

\author{Yuan Cao}
\orcid{0009-0003-6970-1833}
\affiliation{%
  \institution{Key Lab of HCST (PKU), MOE; SCS, Peking University}
  \city{Beijing}
  \country{China}
}
\email{cao_yuan21@stu.pku.edu.cn}

\author{Haochuan Lu}
\orcid{}
\affiliation{%
  \institution{Tencent Inc.}
  \city{Shenzhen}
  \country{China}
}
\email{hudsonhclu@tencent.com}

\author{Hengyu Zhang}
\orcid{}
\affiliation{%
  \institution{Tencent Inc.}
  \city{Shenzhen}
  \country{China}
}
\email{lockerzhang@tencent.com}

\author{Xia Zeng}
\orcid{}
\affiliation{%
  \institution{Tencent Inc.}
  \city{Shenzhen}
  \country{China}
}
\email{xiazeng@tencent.com}

\author{Gang Cao}
\orcid{}
\affiliation{%
  \institution{Tencent Inc.}
  \city{Shenzhen}
  \country{China}
}
\email{marvincao@tencent.com}

\author{Liangchao Yao}
\orcid{}
\affiliation{%
  \institution{Tencent Inc.}
  \city{Shenzhen}
  \country{China}
}
\email{clarkyao@tencent.com}

\author{Yuetang Deng}
\orcid{}
\affiliation{%
  \institution{Tencent Inc.}
  \city{Shenzhen}
  \country{China}
}
\email{yuetangdeng@tencent.com}

\author{Wei Yang}
\orcid{0000-0002-5338-7347}
\affiliation{%
  \institution{University of Texas at Dallas}
  \city{Richardson}
  \country{USA}
}
\email{wei.yang@utdallas.edu}

\author{Tao Xie}
\orcid{0000-0002-6731-216X}
\affiliation{%
  \institution{Key Lab of HCST (PKU), MOE; SCS, Peking University}
  \city{Beijing}
  \country{China}
}
\email{taoxie@pku.edu.cn}

\begin{abstract}
While Large Language Model (LLM) agents show great potential for automated UI navigation such as automated UI testing and AI assistants, their efficiency has been largely overlooked. 
Our motivating study reveals that inefficient UI representation creates a critical performance bottleneck, with UI representation consuming 80 to 99\% of total agent-token costs.
However, UI representation optimization, formulated as the task of automatically generating programs that transform UI representations, faces two unique challenges.
First, the lack of Boolean oracles, which traditional program synthesis uses to decisively validate semantic correctness, poses a fundamental challenge to co-optimization of token efficiency and completeness. Second, the need to process large, complex UI trees as input while generating long, compositional transformation programs, making the search space vast and error-prone. 

Toward addressing the preceding limitations, we present \toolname{}, the first automated optimization framework that synthesizes UI transformation programs by conducting constraint-based optimization with structured decomposition of the complex synthesis task. First, \toolname{} restricts the program space using a domain-specific language (DSL) that captures UI-specific operations, enforcing program validity and reducing search space by construction. 
Second, \toolname{} conducts LLM-based iterative refinement with correctness and efficiency rewards, providing structured guidance for achieving the desired efficiency-completeness co-optimization. 
\toolname{} operates as a lightweight plugin that applies transformation programs for seamless integration with existing LLM agents, requiring minimal modifications to their core logic.
Evaluations across three UI navigation benchmarks spanning Android and Web platforms with five LLMs demonstrate that \toolname{} achieves 48.7\% to 55.8\% token reduction with minimal runtime overhead while maintaining or improving agent performance. Real-world industry deployment at WeChat further validates the practical impact of \toolname{}.
\end{abstract}

\maketitle

\section{Introduction}
Recent advances in Large Language Models (LLMs)~\cite{openai2023gpt4,yang2024qwen2,liu2024deepseek,guo2025deepseek} have enabled the development of LLM-powered Graphical User Interface (GUI) agents that demonstrate promising capabilities in automated software interaction tasks including automated UI testing~\cite{ran2024guardian, feng2024enabling, liu2024make, liu2025guipilot}, intelligent user assistance~\cite{gao2023assistgui, hong2024cogagent, qin2025ui, hu2024dawn}, and application vulnerability detection~\cite{kouliaridis2024assessing, keltek2025lsast, hu2023large, lu2024grace}. 
LLM agents operate by receiving task instructions and iteratively interacting with applications.
These agents capture the current UI state through structured representations (e.g., Android accessibility trees or web DOM trees), reason over the content to plan actions, execute interactions, and repeat until the task is completed.
This automated navigation enables promising applications for software quality assurance and user-experience optimization.
However, as these agents transition from research prototypes to production deployments, the efficiency challenges hinder their practical deployment.

The adoption of LLM agents in increasingly complex UI environments has exposed inefficient UI representations as a critical bottleneck, as shown in our study (Section~\ref{sec::prelim}).
UI representations account for over 80\% of total token usage across studied agent frameworks\cite{zhang2023appagent, wang2024mobilev2, yao2023react} and benchmarks\cite{li2024effectsdatascaleuiandroidcontrol, ran2025beyond, deng2023mind2webgeneralistagentweb}, creating substantial computational overhead, and limits scalability, particularly for advanced agent designs and complex industrial applications that require numerous interaction rounds to accomplish tasks.

Despite the importance of transforming UI representations for LLM efficiency, existing solutions introduce new challenges in terms of token efficiency and semantic completeness. 
First, LLM-based transformation approaches~\cite{deng2023mind2webgeneralistagentweb,furuta2023multimodal,gur2023real} would paradoxically increase token consumption because the raw UI hierarchy must first be processed by the model before any optimization occurs. Although lightweight models have been explored~\cite{gur2023real, furuta2023multimodal}, such approaches suffer from poor generalization across different UI contexts and platforms, limiting their practicality.
Second, heuristic-based transformation approaches can lose critical information such as hierarchical information between nodes for LLM reasoning. LLM4Mobile~\cite{wang2023enabling} extracts all leaf nodes in the UI tree to simplify the complexity, but this flattening approach discards parent-child relationships that are essential for understanding UI structure and navigation paths. Similarly, filtering-based heuristics that remove ``irrelevant'' elements~\cite{droidGPT,wen2024autodroid,vu2024gptvoicetasker} risk eliminating contextually important information that may be crucial for specific tasks. These approaches reduce tokens but often compromise semantic completeness, degrading agent performance when critical UI context is lost. 
% The fundamental challenge lies in achieving an optimal balance between token efficiency and semantic preservation—a problem that requires principled optimization rather than ad-hoc heuristics.

To fundamentally tackle the preceding limitations of existing work, we introduce UI representation optimization as a program synthesis problem, termed as \textit{UI-PS}. Specifically, UI-PS aims to automatically synthesize transformation programs that convert raw UI representations (e.g., DOM trees, accessibility trees) into optimized versions that preserve essential semantic information for LLM perception while significantly reducing token consumption. Unlike manual optimization approaches that require extensive domain expertise, this automated synthesis framework adapts to diverse UI structures and scales across applications, enabling efficient and practical agent deployment in real-world settings.

While the formulation of UI-PS provides a principled framework, solving the problem effectively faces two major challenges.

\noindent\textbf{Challenges.} \textit{Co-optimization for Efficiency and Completeness.} Unlike traditional program synthesis tasks that aim to generate programs with definitive, verifiable outputs—such as producing a program that satisfies a set of input-output examples or conforms to a formal specification—UI-PS tackles an inherently different goal: solving a multi-objective optimization problem. Specifically, UI-PS seeks to reduce the token length of UI representations to improve LLM efficiency while preserving the completeness of semantic information required for downstream perception tasks. The multi-objective optimization of UI-PS lacks the clear Boolean oracles~\cite{solar2008program, alur2013syntax} or definitive success criteria~\cite{gulwani2011automating, li2022competition} (such as test-case assertions or satisfiability checks) that both specification-based and example-based synthesis approaches~\cite{gulwani2017program} depend upon. As a result, standard specification-based and example-driven synthesis approaches fall short when applied to UI-PS, which requires reasoning under soft objectives and partial credit signals rather than binary correctness.

\textit{Combinatorial explosion of the program search space.}  Given the complexity and size of UI inputs—often represented as deeply nested trees with dozens or hundreds of nodes, code models (or any other program synthesis approaches) must reason over the entire structure and produce a long, coherent transformation program that applies globally. The synthesis process involves not only selecting which parts of the UI to simplify but also generating precise transformation logic that composes correctly across multiple levels of the hierarchy. Without explicit structural guidance, models tend to either produce trivial programs that fail to simplify the UI meaningfully or generate aggressive transformations that break semantic behavior. Furthermore, even minor faults in a single part of the generated program can lead to cascading failures, making end-to-end generation brittle and difficult to verify. As a result, the vast, unstructured search space and the absence of intermediate validation signals pose significant obstacles to effective code generation in this setting.

While UI trees may seem too complex for heuristics or direct LLM-based transformation, this complexity can be managed due to two key properties of UI development and rendering. 
First, much of the complexity stems from implementation choices where multiple nodes or layers implement a single user-perceived component. Developers deliberately adopt nested hierarchies for modularity and performance optimization, introducing fragmented layers for styling, layout, and event handling.  As shown in Figure~\ref{fig::motiv}, the search component maintains its own internal node hierarchy for placeholder text, icon positioning, and input handling, while the input field further contains separate nodes for the editable text area, selection highlighting, cursor positioning, and touch-event handling—many of which are structural rather than semantic. 
Second, UI complexity is cumulative, arising from the composition of such redundant sub-structures throughout the hierarchy. This cumulative nature lends itself to  a divide-and-conquer approach that decomposes the optimization problem into manageable local transformations. Rather than attempting to optimize the entire tree globally, we identify and eliminate redundant intermediate nodes through local reduction operations. By recursively merging non-semantic nodes from the bottom up, we preserve semantic completeness while improving representation efficiency.

To instantiate the preceding insights, we propose \toolname{}, an automated optimization framework for synthesizing UI transformation programs with two major strategies, as shown in Figure~\ref{fig::overview}.
First, \toolname{} restricts the transformation space using a domain-specific language (\textit{DSL}).
The DSL captures UI-specific merge operations, enforcing semantic completeness by targeting the fragmentation patterns while making the synthesis problem friendly for code LLMs.
Second, we design an LLM-based iterative refinement process that monotonically improves the quality of synthesized programs.
In this process, we use evaluators to check whether the previously synthesized transformation programs violate the semantic completeness and evaluate the efficiency of reducing token usage. Using the additional information that can be obtained locally, we prompt the code LLM to refine program synthesis until the optimization efficacy reaches the desired level.

To easily integrate the preceding synthesized programs for improving the efficiency of any LLM agent, \toolname{} operates the UI transformation as a plugin that intercepts UI representations during agent runtime.
As shown in Figure~\ref{fig::overview} (b), \toolname{} captures screen states (accessibility trees or DOM trees), applies transformation programs, and provides optimized representations to agents, without requiring modifications to their core logic.

To evaluate the effectiveness, generalization, and practicality of \toolname{}, we conduct extensive experiments across three diverse datasets spanning Web and Android UI navigation benchmarks with five LLMs, complemented by real-world industry deployment at WeChat. 
Our evaluation results highlight three key findings. 
First, \toolname{} substantially improves LLM efficiency while enhancing rather than compromising agent performance, achieving an average reduction of 48.7\% to 55.8\% in UI representation tokens while maintaining semantic completeness.
The improvement is general and consistent across different LLMs, agent designs, and UI platforms.
Second, our ablation studies confirm the necessity of the DSL.
While state-of-the-art proprietary LLMs can synthesize seemingly plausible UI transformation programs, the programs either sacrifice semantic completeness or perform superficial simplification.
Third, industry deployment experiences at WeChat show that \toolname{} improves 35\% query per minute (QPM) and reduces 26.1\% latency in production environments, validating the practical impact.

In summary, this paper makes the following major contributions:

\begin{itemize}
     \item A motivating study examining the impact of UI representations on LLM agent efficiency. Our results show opportunities for improving UI representation efficiency and highlight the necessity of automated optimization approaches.
    
    \item Design and implementation of \toolname{}, an automated optimization framework that iteratively refines transformation programs with a DSL to achieve co-optimization of token efficiency and semantic completeness. \toolname{} can be integrated with existing agent frameworks and UI platforms with minimal modification.

    \item Comprehensive evaluations across multiple benchmarks, LLMs, agents, and real-world industry deployment at WeChat demonstrating \toolname{}'s effectiveness and generalization in improving LLM agent efficiency while even improving agent performance.
\end{itemize}

\section{Motivating Study}\label{sec::prelim}
LLM agents have demonstrated remarkable capabilities in automating complex tasks across diverse user interface environments, from web navigation to mobile app interaction. Existing work has been focusing on designing sophisticated agent architectures, reasoning strategies, and task decomposition approaches to improve agent performance and reliability~\cite{yao2023react,wang2024mobilev2,zhang2023appagent}. 
Although the effectiveness of LLM agents is critical, the efficiency of UI representations has received limited attention despite its fundamental importance for practical deployment. To investigate the impact of UI representations on LLM agent efficiency, we conduct a motivating study that examines token consumption patterns across different agent frameworks. Our analysis reveals that UI representation inefficiency poses a significant bottleneck that limits the scalability and real-world applicability of LLM agents, motivating the need for principled optimization approaches.

\subsection{Study Setup}\label{sec::token-analysis}
To quantify token consumption caused by UI representations, we conduct a comprehensive measurement study of popular LLM agents on existing UI navigation benchmarks.

\noindent\textbf{Study subjects.} We analyze token consumption of three popular LLM agents: ReAct~\cite{yao2023react}, Mobile-Agent-V2~\cite{wang2024mobilev2}, and AppAgent~\cite{zhang2023appagent}. These agents are evaluated on AndroidControl~\cite{li2024effectsdatascaleuiandroidcontrol}, a widely-used Android UI navigation benchmark, and Mind2Web~\cite{deng2023mind2webgeneralistagentweb}, a comprehensive web navigation benchmark. Details of the LLMs and benchmarks can be found in Section~\ref{sec::eval::setup}. We execute the agents on both benchmarks and systematically log all prompts used for LLM invocations to capture real-world token usage patterns.

\noindent\textbf{Token consumption measurement.} We categorize prompts into six functional components: (1) \textbf{system instruction} describing the agent's role and capabilities, (2) \textbf{action space} defining available actions and their parameters, (3) \textbf{task description} specifying the current objective, (4) \textbf{UI representation} containing UI state information such as simplified HTML and accessibility trees, (5) \textbf{history context} including historical actions and navigation context, and (6) \textbf{output format} providing response format specifications and examples. This categorization captures the design of all studied LLM agents across different frameworks.

Token consumption is measured using the respective tokenizers of GPT-4o~\cite{openai2023gpt4}, DeepSeek-V3~\cite{liu2024deepseek}, and Qwen-2.5~\cite{yang2024qwen2} to reflect real-world deployment scenarios. We encode each prompt component using the corresponding tokenizer's encode function and sum the token counts across all components to obtain the total token consumption per UI interaction. The remaining analysis focuses on identifying which components contribute most significantly to overall computational costs.

\begin{table*}[htbp]
\centering
\caption{Token consumption breakdown by model, benchmark, and platform. For each configuration, the average number of tokens per UI page is reported according to six prompt components, which are system instruction (\textbf{System}), action space (\textbf{Action}), task description (\textbf{Task}), UI representation (\textbf{UI}), action history (\textbf{Context}), and output format (\textbf{Format}).}
\label{tab:token_usage_breakdown}
\resizebox{\textwidth}{!}{
\begin{tabular}{lllcccccccc}
\toprule
 \textbf{Model} & \textbf{Benchmark} & \textbf{Agent} & \textbf{System} & \textbf{Action} & \textbf{Task} & \textbf{UI} & \textbf{Context} & \textbf{Format} & \textbf{Total} & \textbf{UI Token Ratio}\\
\midrule
GPT-4o~\cite{openai2023gpt4} & AndroidControl~\cite{li2024effectsdatascaleuiandroidcontrol} & ReAct~\cite{yao2023react} &  98 & 158 & 11 & 2,793 & 16 & 144 & 3,220 & 86.7\% \\
GPT-4o~\cite{openai2023gpt4} & AndroidControl~\cite{li2024effectsdatascaleuiandroidcontrol} & AppAgent~\cite{zhang2023appagent} & 195 & 286 & 11 & 2,793 & 16 & 187 & 3,488 & 80.1\% \\
GPT-4o~\cite{openai2023gpt4} & AndroidControl~\cite{li2024effectsdatascaleuiandroidcontrol} & Mobile-Agent-V2~\cite{wang2024mobilev2} &  287 & 231 & 11 & 2,793 & 16 & 162 & 3,500 & 79.8\% \\
DeepSeek-V3~\cite{liu2024deepseek} & AndroidControl~\cite{li2024effectsdatascaleuiandroidcontrol} & ReAct~\cite{yao2023react} &  122 & 190 & 13 & 2,951 & 18 & 158 & 3,452 & 85.5\% \\
Qwen-2.5~\cite{yang2024qwen2} & AndroidControl~\cite{li2024effectsdatascaleuiandroidcontrol} & ReAct~\cite{yao2023react} &  99 & 158 & 11 & 2,751 & 16 & 146 & 3,181 & 86.5\% \\
Qwen-2.5~\cite{yang2024qwen2} & Mind2Web~\cite{deng2023mind2webgeneralistagentweb} & ReAct~\cite{yao2023react} & 98 & 186 & 28 & 51,648 & 40 & 146 & 52,146 & 99.0\% \\
\bottomrule
\end{tabular}
}
\end{table*}

\subsection{Study Results}  
Table~\ref{tab:token_usage_breakdown} shows the breakdown of token consumption by different prompt components across various LLM models, benchmarks, and agent frameworks. Specifically, we measure the token counts for six types of prompt components: system instruction, action space, task description, UI representation, action history context, and output format. The total token consumption per UI page is then calculated by summing all component contributions.
As shown in Table~\ref{tab:token_usage_breakdown}, UI representations consistently dominate token consumption across all configurations. By examining results from all model-benchmark combinations, we observe that UI representations account for 80-99\% of total tokens, ranging from 2,751 tokens on AndroidControl to 51,648 tokens on Mind2Web. In contrast, other prompt components remain relatively stable, with system instruction (98-122 tokens), action space (158-190 tokens), task description (11-28 tokens), context (16-40 tokens), and format (144-158 tokens) contributing minimally to overall consumption.
Our findings suggest that focusing on UI representation optimization has significant potential for reducing computational costs in LLM agent deployment, as this component constitutes the overwhelming majority of token usage across all studied configurations.

\section{Motivating Example}\label{sec::motiv}

\begin{figure}[t]
    \centering
    \includegraphics[width=\linewidth]{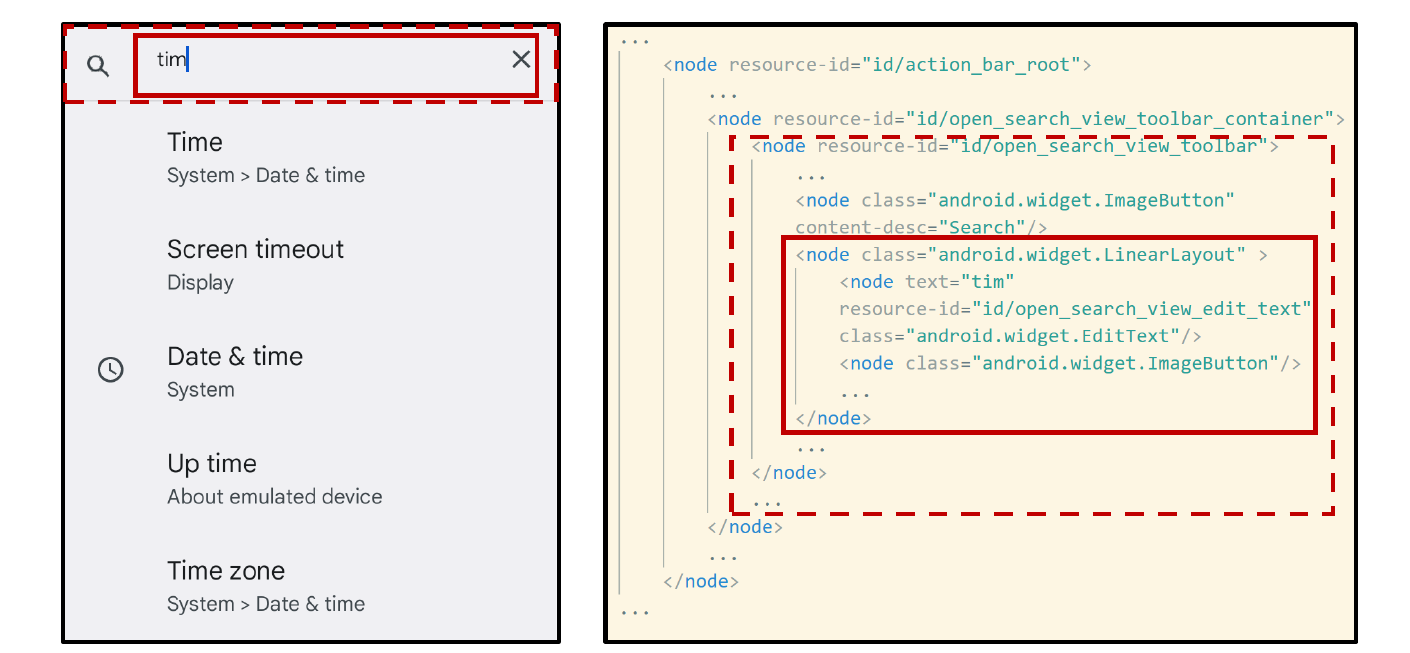}
    \caption{A simplified example of UI Tree (Right) and its corresponding screenshot (Left).}
    \Description{A diagram showing a UI tree on the right and a corresponding screenshot on the left.}
    \label{fig::motiv}
\end{figure}

Modern mobile applications present complex user interfaces that appear simple to users but are built from intricate hierarchical structures. Figure~\ref{fig::motiv} presents a straightforward search UI on Android. 
Yet beneath this visual simplicity lies a complex UI tree representation.

\textbf{UI trees} represent interface structure through nested XML nodes, where each node contains properties such as type, text, position, and interaction capabilities. However, what appears as a single functional component to users often requires multiple implementation nodes. For example, the search component in Figure~\ref{fig::motiv} internally maintains separate nodes for container layout and positioning, placeholder text rendering, input field boundaries and styling, icon positioning and touch handling, and background rendering and visual effects.
The \textbf{implementation fragmentation} serves important engineering purposes such as modularity, maintainability, and performance optimization, but creates \textbf{semantic redundancy} from an LLM perspective. A single conceptual search box becomes 8-12 nodes, dramatically inflating token usage while providing little additional semantic value.

A natural approach would be to optimize UI representations through unconstrained program synthesis, automatically removing redundant nodes while preserving semantic information. However, doing so leads to two critical challenges.
First, the combinatorial space of possible tree transformations grows exponentially with tree size. For a tree with $n$ nodes, there are $2^n$ possible node removal combinations, making exhaustive search intractable even for moderate-sized interfaces.
Second, without principled constraints, optimization heuristics may inadvertently remove semantically important nodes, breaking the structural coherence needed for LLM comprehension. The challenge is distinguishing between implementation artifacts (safe to remove) and semantic content (must preserve).

Our key insight is that UI fragmentation follows predictable structural patterns driven by recursive locality. UI trees are constructed through hierarchical composition, where parent nodes aggregate child functionality—a decomposition process driven by developer implementation needs rather than user perception. We can reverse this process through \textit{local reduction operations} that work bottom-up from leaf nodes (inverse to the top-down rendering process): recursively merging nodes that serve purely structural purposes, preserving semantic boundaries where users perceive distinct functional components, and maintaining verifiability through local parent-child relationships. This approach transforms an intractable global optimization problem—reasoning about exponential tree transformations—into a series of \textbf{local and verifiable decisions}. By constraining synthesis to simple parent-child merge operations that are both \textbf{safe} (preserving semantics) and \textbf{efficient} (reducing tokens), we enable principled UI consolidation that scales reliably across diverse applications, motivating the design of \toolname{}.

\section{Design of \toolname{}}\label{sec::approach}
We present \toolname{}, an optimization framework that iteratively synthesizes UI-transformation programs within a carefully designed DSL to improve the token efficiency of UI representations while preserving the completeness of UI semantics for LLM agents.

\begin{figure*}[t]
    \centering
    \includegraphics[width=0.9\linewidth]{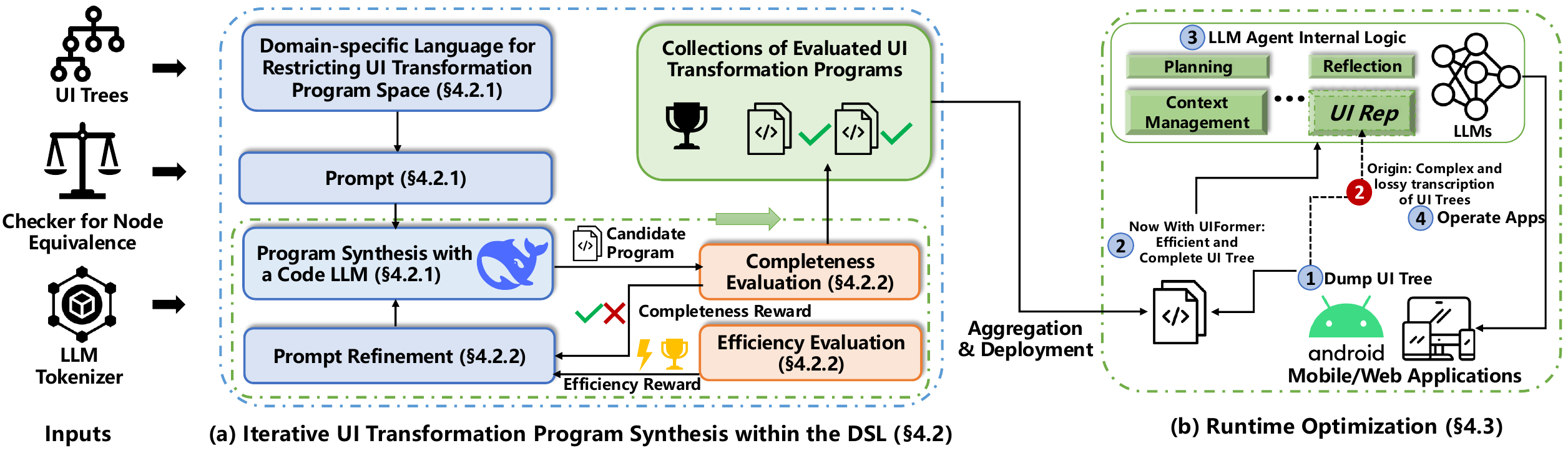}
    \caption{Overview of \toolname{}. \toolname{} operates in two phases: \textit{(a) offline program synthesis} to iteratively refine UI transformation programs within a DSL, and \textit{(b) runtime optimization} to seamlessly integrate the synthesized programs with existing LLM agents.}
    \Description{A schematic diagram illustrating the two-phase workflow of \toolname{}. 
    Phase (a) shows the offline program synthesis, where a domain-specific language guides iterative refinement of UI transformation programs. 
    Phase (b) shows runtime optimization, where the synthesized programs act as a lightweight plugin to existing LLM agents.}
    \label{fig::overview}
\end{figure*}

\subsection{Overview of \toolname{}}
Figure~\ref{fig::overview} presents the overview of \toolname{}, which operates in two distinct phases: an offline program-synthesis phase (Section~\ref{sec::approach::offline}) and a runtime-optimization phase (Section~\ref{sec::approach::online}).

During the \textbf{offline phase}, to co-optimize for token efficiency and semantic completeness while avoiding combinatorial explosion of program search space, as shown in Figure~\ref{fig::overview} (a), we design a DSL (detailed in Section~\ref{sec::approach::offline::dsl}) and an iterative refinement process (detailed in Section~\ref{sec::approach::offline::iteration}) over the DSL and local evaluations to automatically improve the synthesized programs.
The DSL provides expressive primitives for UI transformation operations while restricting the operations to occur between adjacent nodes in the UI tree, reducing the task complexity for the code LLM.
The iterative refinement process bootstraps the code LLM with token reduction reward and semantic completeness reward to evolve increasingly sophisticated transformation programs without requiring extensive annotations.

During the \textbf{runtime phase}, to integrate with existing LLM agents seamlessly, we implement \toolname{} as an agent-agnostic plugin requiring minimal modification to LLM agents. As shown in Figure~\ref{fig::overview} (b), \toolname{} transparently intercepts UI representations before they reach LLM agents, applies the synthesized transformations, and delivers optimized UI representations that maintain semantic completeness while improving token efficiency.

\begin{algorithm}[t]
\caption{DSL-Restricted Program with Iterative Refinement}
\label{alg:program_synthesis}
\begin{algorithmic}[1]
\Require UI training examples $\mathcal{E} = \{(T_i^{\text{orig}}, T_i^{\text{target}})\}_{i=1}^{N}$
\Ensure Verified program library $\mathcal{L}$

\State Initialize $\mathcal{L} \leftarrow \emptyset$, feedback history $\mathcal{F} \leftarrow \emptyset$

\While{not converged \textbf{and} $\text{iter} < \text{MAX\_ITER}$}
    \State \textcolor{blue}{\textit{// Generate DSL-constrained candidate programs}}
    \State $\mathcal{P} \leftarrow \text{LLM\_Synthesize}(\text{DSL\_Template}, \mathcal{E}, \mathcal{F})$
    
    \For{each program $p \in \mathcal{P}$}
        \State \textcolor{blue}{\textit{// Evaluate completeness and efficiency}}
        \State $R_{\text{total}} \leftarrow \sum_{(T^{\text{orig}}, T^{\text{target}}) \in \mathcal{E}} R(p(T^{\text{orig}}), T^{\text{target}})$
        
        \If{$R_{\text{total}} \geq \text{THRESHOLD}$}
            \State $\mathcal{L} \leftarrow \mathcal{L} \cup \{p\}$ \textcolor[rgb]{0,0.5,0}{\textit{// Accept verified program}}
        \Else
            \State $\mathcal{F} \leftarrow \mathcal{F} \cup \{\text{GenerateFeedback}(p, \mathcal{E})\}$ \textcolor{red}{\textit{// Refine with efficiency and completeness reward}}
        \EndIf
    \EndFor
\EndWhile

\State \Return $\mathcal{L}$ \textcolor{blue}{\textit{// Library of verified transformation programs}}
\end{algorithmic}
\end{algorithm}

\subsection{DSL-restricted Program Synthesis with Iterative Refinement}\label{sec::approach::offline}

The offline program synthesis phase takes UI transformation examples as input and generates a library of verified programs that preserve semantic completeness while maximizing token efficiency. As shown in Algorithm~\ref{alg:program_synthesis}, we employ iterative refinement where LLMs generate DSL-constrained candidate programs, evaluate them against training examples using our composite reward function, and iteratively improve through structured feedback. Programs that achieve sufficient completeness and efficiency scores are added to the verified library $\mathcal{L}$, while failed attempts generate specific feedback to guide subsequent synthesis iterations. This process continues until convergence, producing a robust set of transformation programs that generalize across diverse UI trees.

\subsubsection{DSL Design for UI Transformation Programs.}\label{sec::approach::offline::dsl}

General-purpose programming languages create an intractably large search space for UI transformation tasks due to their unrestricted operation sets. Additionally, unconstrained program generation produces transformations that are difficult to verify and often fail to generalize across different UI structures.

To address these challenges, we constrain the program synthesis space through a domain-specific language (DSL) that decomposes the optimization problem into local, verifiable operations. 
Instead of synthesizing transformations over entire UI trees, our DSL restricts LLMs to generate recursive code snippets that operate on parent-child node pairs.

\begin{figure}
    \centering
\begin{lstlisting}[
    language=Python,
    basicstyle=\small\ttfamily,
    commentstyle=\color{gray},
    keywordstyle=\color{blue},
    numberstyle=\tiny,
    numbers=left,
    frame=single,
    breaklines=true,
    escapeinside={(*}{*)}
]
def transform_node(node, child_views):
    """
    Input: UINode, List[View]
    Output: List[View]
    """
    # LEAF NODE HANDLING
    if not child_views:  # Leaf node case
        if (*\textit{<leaf\_filter\_condition>}*):
            return []
        else:
            leaf_view = CREATE_VIEW(node, (*\textit{<leaf\_properties>}*))
            return [leaf_view]
    # FILTERING
    if FILTER((*\textit{<binary\_predicate\_expression>}*)):
        return child_views
    #MERGING DECISION
    if (*\textit{<merge\_condition\_expression>}*):
        merged_view = MERGE_OPERATION(node, child_views)
        return [merged_view]
    # PASS-THROUGH
    return child_views
\end{lstlisting}
    \caption{DSL Specifications for Code LLMs.}
    \Description{DSL Specifications for Code LLMs.}
    \label{fig:dsl-template}
\end{figure}

\noindent\textbf{DSL Signature.} As shown in Figure~\ref{fig:dsl-template}, each synthesized program $P$ transforms a parent UI node $n$ and its processed child views $C = \{c_1, c_2, \ldots, c_k\}$ into a refined view set $V'$.

\noindent\textbf{Core Operations.} The DSL provides three operations that capture common UI optimization patterns. First, \textbf{Filtering} selectively removes UI elements based on semantic relevance, reducing token overhead from irrelevant UI nodes. 
Second, \textbf{Merging} combines related child views into consolidated representations when semantic boundaries permit aggregation. Third, \textbf{Pass-through} preserves existing view boundaries when no optimization is beneficial, maintaining semantic integrity.
For leaf nodes, the DSL enables conditional view creation or filtering based on node properties, ensuring fine-grained control over information retention.

\noindent\textbf{Synthesis Target.} The LLM synthesizes the conditional logic within this template, enabling adaptive UI optimization across diverse interface patterns. \textit{Condition expressions} include leaf filter conditions that determine when to discard irrelevant elements (e.g., decorative dividers, empty containers), binary predicate expressions for parent node filtering based on structural properties (e.g., node type, attribute patterns, child count), and merge condition expressions that identify semantically related components suitable for consolidation (e.g., consecutive text elements, grouped form inputs). 
\textit{Property extraction rules} specify which semantic attributes to preserve during transformation, such as textual content, interaction affordances, spatial relationships, and accessibility labels. 
These rules ensure that critical UI information is retained while eliminating redundant structural details. 
This constrained synthesis space enables efficient search while maintaining expressiveness for diverse UI patterns. 
The recursive application of these local transformations yields globally optimized UI representations that preserve semantic completeness while reducing token usage.

\begin{figure}[t]
    \centering
    \includegraphics[width=0.75\linewidth]{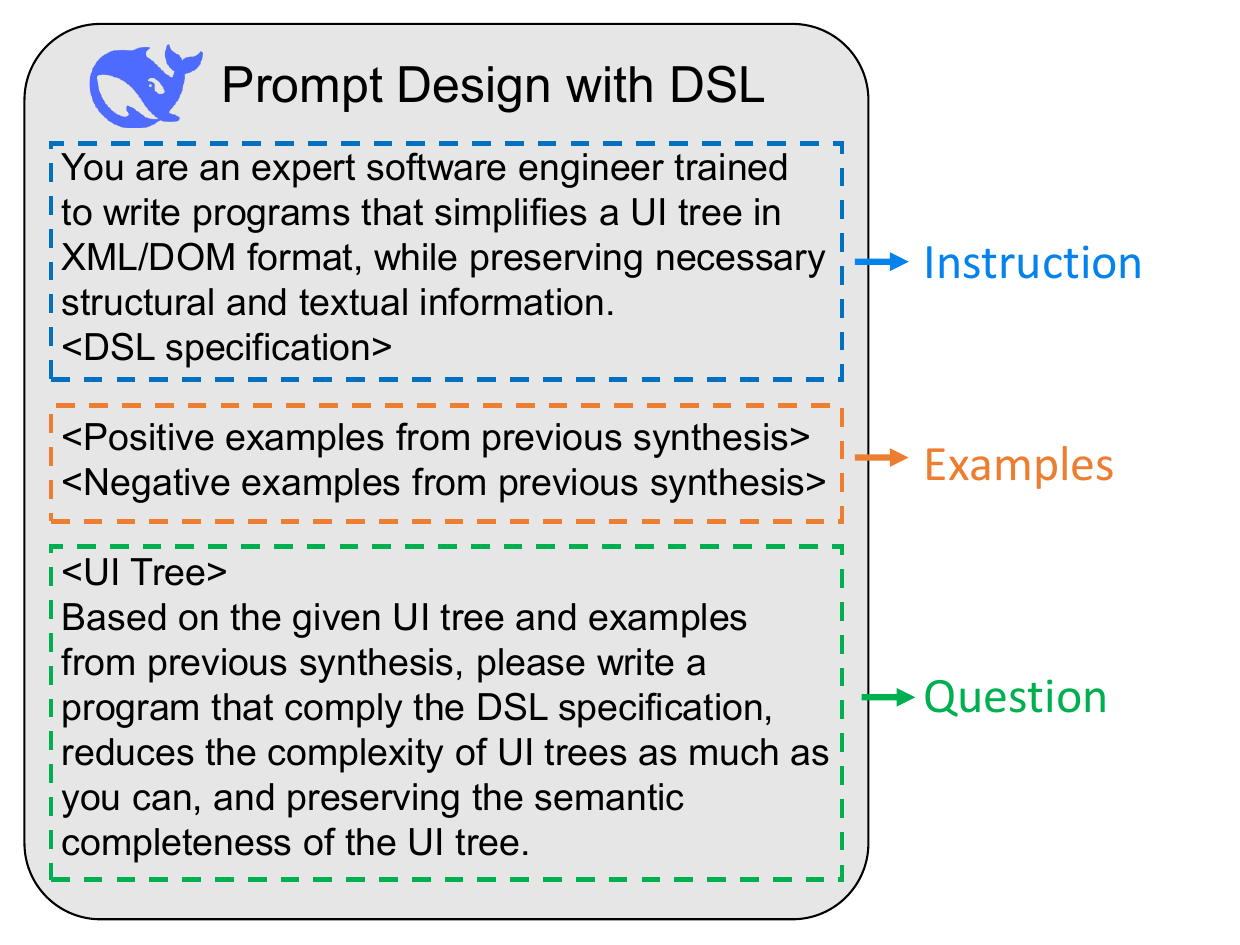}
    \caption{Prompt for the Code LLM.}
    \Description{The figure shows an example input prompt provided to a code-oriented LLM, illustrating the format and content of the instructions.}
    \label{fig::code_synthesis_prompt}
\end{figure}

\subsubsection{Iterative Refinement with Efficiency and Completeness Feedback.}\label{sec::approach::offline::iteration}

Single-shot program synthesis rarely produces optimal solutions for complex UI transformation tasks, as LLMs lack visibility into transformation failures and cannot incorporate evaluation feedback. Consequently, \toolname{} designs a feedback-driven iterative refinement loop.
As shown in Algorithm~\ref{alg:program_synthesis}, we prompt a code LLM with DSL syntax, semantics, and UI transformation examples (from previous iterations) to generate candidate programs through few-shot prompting with carefully curated examples that demonstrate effective transformation patterns as well as transformation failures.

\noindent\textbf{Evaluation and feedback generation.} Each candidate program is immediately evaluated against training examples using a composite reward function:
\begin{equation}
R(p) = R_{\text{completeness}}(p) + R_{\text{efficiency}}(p)
\end{equation}
The completeness component $R_{\text{completeness}}(p)$ assigns large negative penalties (-10) when programs incorrectly merge semantically distinct elements or lose critical UI information, ensuring semantic preservation. The efficiency component $0\leq R_{\text{efficiency}}(p)\leq1$ measures token reduction proportion achieved after applying $p$.

Programs achieving rewards above the acceptance threshold are added to the verified library $\mathcal{L}$, while failing programs contribute to feedback generation. \toolname{} constructs negative examples from transformation programs achieving negative rewards and positive examples from programs achieving the highest rewards without breaking semantic completeness. These concrete examples of successful and failed transformations enrich the prompt shown in Figure~\ref{fig::code_synthesis_prompt} (specifically, examples) for the next iteration, enabling the LLM to systematically learn from previous failures and improve toward high-quality transformation programs that optimize token efficiency and preserving semantic completeness.

\subsection{Runtime UI Optimization}\label{sec::approach::online}
When LLM agents request UI information from UI platforms (e.g., web browsers or Android devices), \toolname{} applies the synthesized transformation programs to optimize UI representations with minimal latency overhead. We design \toolname{} as a lightweight plugin that intercepts and transforms UI content between LLM agents and UI applications, enabling seamless integration with existing agent frameworks while providing universal optimization benefits across diverse agent architectures and LLM models. 
To integrate with \toolname{}, agent developers only need to redirect their UI representation module while leaving all other agent components unchanged. \toolname{} implements a three-stage transformation pipeline that operates within the plugin layer. First, when agents request UI information, the system captures the XML or DOM tree from platform-specific APIs (e.g., web page DOMs, mobile app view hierarchies, desktop accessibility trees).
Second, \toolname{} sequentially applies the synthesized transformation programs to the captured UI tree to obtain a simplified UI tree while preserving the original UI semantics.
Third, \toolname{} constructs optimized UI prompts based on the simplified UI tree.

\section{Evaluation}\label{sec::eval}
In this section, we conduct comprehensive evaluations to answer the following research questions:

\begin{itemize}
    \item \textbf{RQ1 (Efficiency Improvement):} How effective is \toolname{} in improving the efficiency of LLM agents while preserving semantic completeness?
    
    \item \textbf{RQ2 (Ablation Study):} How does each individual component contribute to the effectiveness of \toolname{}?
    
    \item \textbf{RQ3 (Practicality):} How effective is \toolname{} in improving the efficiency of LLM agent service in real-world industry deployment?
\end{itemize}

\subsection{Evaluation Setup}\label{sec::eval::setup}

\subsubsection{Benchmarks and Metrics}

We evaluate \toolname{} on three benchmarks across Android and web UI navigation tasks, measuring both token efficiency and semantic completeness.

\noindent\textbf{Android UI Navigation.} We use two Android benchmarks: \textit{Sphinx}~\cite{ran2025beyond} and \textit{AndroidControl}~\cite{li2024effectsdatascaleuiandroidcontrol}. Sphinx is a challenging mobile navigation benchmark with 458 tasks from 100 popular apps; we use Sphinx-Lite (163 tasks) to avoid login requirements. 
Sphinx evaluates the performance of LLM agents by success rate (SR, the proportion of completed tasks) and average completion proportion (ACP, the average proportion of target functionalities achieved).
AndroidControl contains 14,548 tasks across 833 apps; we sample 500 action steps following prior work~\cite{gou2024navigating}. AndroidControl evaluates the performance of LLM agents by step success rate (Step SR), measuring the proportion of correctly executed individual actions.

\noindent\textbf{Web UI Navigation.} We use Mind2Web~\cite{deng2023mind2webgeneralistagentweb}, a comprehensive web navigation benchmark with tasks across diverse websites. Due to limited budget, we use 50 tasks with top-5 candidate actions for evaluation. Mind2Web evaluates the performance of LLM agents by element classification accuracy (Ele.Acc), operation prediction F1 (Op.F1), and step-level success rate (Step SR).

\noindent\textbf{Evaluation Metrics.} We measure \textit{token efficiency} as the average input tokens per navigation step and \textit{semantic completeness} using benchmark-specific performance metrics (SR, ACP, Step SR, Task SR, Ele.Acc, Op.F1, and Step SR).

\subsubsection{Baselines.}\label{sec::setup::baseline_rep}
We compare LLM's performance with simplified UI representations used by three state-of-the-art approaches~\cite{wang2023enabling,wen2024autodroid,vu2024gptvoicetasker}, with manual heuristics for preserving the hierarchy of UI trees or simplifying UI contents.

\begin{itemize}
\item \textbf{Operation-based UI Representation (Ops)}~\cite{wen2024autodroid} prunes the UI tree by discarding invisible and decorative nodes, and keeps only interactive nodes. It also merges nodes leading to the same screen state to simplify the representation. This preserves the operation structure while reducing irrelevant content.

\item \textbf{Leaf-based UI Representation (Leaf)}~\cite{wang2023enabling} preserves all leaf nodes in the original Android XML UI tree. It converts the tree into HTML format under the assumption that LLMs are well-trained on HTML structures. This representation retains full detail at the leaf level.

\item \textbf{Flattened UI Representation (Flattened)}~\cite{vu2024gptvoicetasker} serializes the UI hierarchy into a flat sequence of semantic elements, categorized by interaction type (e.g., clickable items, text inputs, scrollable views). It summarizes the UI into a high-level description followed by a list of categorized interactive elements.
\end{itemize}

We adopt ReAct~\cite{yao2023react}, a simple yet effective agent framework, which is widely used as the default LLM agents for UI navigation~\cite{wang2023enabling,ran2024guardian,wen2024autodroid,vu2024gptvoicetasker}.
We use the preceding UI representation baselines and \toolname{} to construct the UI representation prompt, while keeping other parts of the prompts the same.
We also use AppAgent~\cite{zhang2023appagent} and Mobile-Agent-v2~\cite{wang2024mobilev2} to evaluate the generalization of \toolname{}.

\subsubsection{Experiment Platforms.} All experiments are conducted on the official Android x64 emulators running Android 12.0 on a server with four
AMD EPYC 7H12 64-Core Processors. Each emulator is allocated
with 4 dedicated CPU cores, 2 GB RAM, and 2 GB internal storage.
We serve local models with 8 NVIDIA H20 GPU Cards using vLLM inference framework~\cite{kwon2023efficient}.
We use DeepSeek-V3~\cite{liu2024deepseek} as the code LLM for program synthesis due to its high performance in code generation and cost-effectiveness.

\subsection{RQ1: Efficiency Improvement}
We evaluate \toolname{}'s effectiveness in reducing token usage while preserving semantic completeness for LLM reasoning across different platforms, models, and agent frameworks.

\subsubsection{Main results.} Table~\ref{table::rq4::method} and ~\ref{tab:mind2web_full} present the efficiency results of different UI representations for the ReAct agent on Android and Web, respectively. 
In addition, we present the results of the efficiency results across different LLMs in Table~\ref{table::rq4::llm}, and Table~\ref{table::rq4::agent} validates effectiveness across agent frameworks.
From these tables, we have four major observations.

\begin{table}[t]
\centering
\caption{Effectiveness of different UI representation approaches on the Sphinx benchmark, evaluated using GPT-4o and the ReAct agent. \textcolor{red}{$\downarrow$} and \textcolor[rgb]{0,0.5,0}{$\uparrow$} indicate lower and higher is better, respectively.}\label{table::rq4::method}
\begin{tabular}{lcccc}
\toprule
\textbf{Approach} & \textbf{Tokens \textcolor[rgb]{1,0,0}{$\downarrow$}}  & \textbf{ACP \textcolor[rgb]{0,0.5,0}{$\uparrow$}} & \textbf{SR \textcolor[rgb]{0,0.5,0}{$\uparrow$} }  \\
\midrule
\textbf{Ops~\cite{wen2024autodroid}}  & 1,206 (\textcolor[rgb]{0,0.5,0}{-51\%}) &  35.12 & 30.67    \\
\textbf{Leaf~\cite{wang2023enabling}} & 1,033 (\textcolor[rgb]{0,0.5,0}{-42\%}) &  32.16 & 27.61    \\
\textbf{Flattened~\cite{vu2024gptvoicetasker}}  & 648 (\textcolor[rgb]{0,0.5,0}{-8\%}) &  19.88 & 15.33    \\
\midrule
\textbf{\toolname{} (Ours)} & \textbf{596}  &\textbf{37.27} & \textbf{34.36}     \\
\bottomrule
\end{tabular}
\end{table}

\noindent\textbf{Co-optimization of token efficiency and agent performance.} \toolname{} achieves the best task performance while using the fewest tokens across all benchmarks. 
Instead of baseline heuristics sacrificing semantic completeness for efficiency, \toolname{} co-optimizes the token efficiency and semantic completeness. While Flattened achieves high token efficiency, it sacrifices the semantic completeness, making it difficult for LLMs to understand overall UI structure.
On Sphinx, \toolname{} improves the agent performance by 3.69\% absolute improvement in success rate over the best baseline while reducing tokens by 51\% compared to the Ops UI representation. 
On Mind2Web, the improvements are even more substantial. \toolname{} achieves 53.71\% step success rate with only 6,102 tokens, outperforming the best baseline by 7.16\% absolute improvement while using 63\% fewer tokens, demonstrating \toolname{}'s effectiveness to co-optimize token efficiency and semantic completeness (i.e., agent performance).

\noindent\textbf{Platform-adaptive optimization.} The results reveal \toolname{}'s ability to automatically adapt to platform-specific requirements. 
On web benchmark Mind2Web, performance improves as token count decreases across baselines, with Flattened (16,683 tokens) outperforming Ops (52,146 tokens), suggesting web UIs benefit from aggressive simplification due to their verbose DOM structures. Conversely, on Android benchmark Sphinx, performance degrades with reduced tokens, as Ops (1,206 tokens) outperforms Flattened (648 tokens), indicating mobile UIs require more complete representations to preserve semantic completeness. 
Despite these opposing platform dynamics, \toolname{} consistently achieves optimal performance on both platforms, demonstrating the importance of automated optimization and the effectiveness of \toolname{}.

% \begin{table}[t]
% \caption{
% Results on \textsc{Mind2Web}. 
% % Metrics include element classification accuracy (Ele.Acc), operation prediction F1 (Op.F1), and step-level success rate (Step SR). Token consumption is reported in average input tokens per step. 
% }
% \label{tab:mind2web_full}
% \centering
% \begin{tabular}{lcccc}
% \toprule
% \textbf{Approach} & \textbf{Tokens} & \textbf{Ele.Acc} & \textbf{Op.F1} & \textbf{Step SR}  \\
% \midrule
% \textbf{Ops~\cite{wen2024autodroid}} & 52,146 & 49.53 & 65.94 & 41.23 \\
% \textbf{Leaf~\cite{wang2023enabling}}  &  20,712 & 53.59 & 66.15 & 44.57   \\
% \textbf{Flattened~\cite{vu2024gptvoicetasker}}  & 16,683 & 55.58 & 70.15 & 46.55   \\
% \midrule
% \textbf{\toolname{} (Ours)} & \textbf{6,102} & \textbf{61.72} & \textbf{72.59}  &  \textbf{53.71} \\
% \bottomrule
% \end{tabular}
% \end{table}

\begin{table}[t]
\caption{
Results on \textsc{Mind2Web}. 
}
\label{tab:mind2web_full}
\centering
\begin{tabular}{lcccc}
\toprule
\textbf{Approach} & \textbf{Tokens} & \textbf{Ele.Acc} & \textbf{Op.F1} & \textbf{Step SR}  \\
\midrule
\textbf{Baseline~\cite{wen2024autodroid}} & 52,146 & 49.53 & 65.94 & 41.23 \\
\midrule
\textbf{\toolname{} (Ours)} & \textbf{6,102 (-88\%)} & \textbf{61.72 (+24\%) } & \textbf{72.59 (+10\%)}  &  \textbf{53.71 (+30\%)} \\
\bottomrule
\end{tabular}
\end{table}

\noindent\textbf{Consistent improvements across diverse LLMs.} Table~\ref{table::rq4::llm} shows that \toolname{} delivers consistent benefits across five different language models. Every model achieves higher success rates with \toolname{} while using approximately 50\% fewer tokens. The improvements range from 0.61\% to 7.36\% absolute improvement, with an average improvement of 4.05\% absolute improvement across all models. The consistency improvement indicates the generalizable effectiveness of \toolname{}.

\noindent\textbf{Framework-agnostic effectiveness.} As shown in Table~\ref{table::rq4::agent}, across three different agent frameworks (ReAct, AppAgent, and Mobile-Agent-v2), \toolname{} consistently reduces token usage by approximately 45\% while improving step success rates by 0.8\% to 2.2\% absolute improvement. The results further confirms the generalizable effectiveness of \toolname{}, benefiting to all LLM agents for efficiency improvement.

\subsubsection{Runtime Overhead.} Since we adopt rule-based UI transformation programs instead of introducing additional models for UI optimization, the runtime overhead is extremely low. To quantify this overhead, we randomly sample 1000 UI trees from the UI traces collected from the experiments on AndroidControl and log the processing time. The processing time ranges from 0.02ms to 80.1ms, with an average of 5.7ms per UI tree, which is almost negligible.

\setlength{\tabcolsep}{6pt}
\begin{table*}[t]
\centering
\caption{Performance comparison of five LLMs with and without \toolname{} on the Sphinx and AndroidControl benchmarks. All evaluations use the ReAct agent and models without \toolname{} directly take the view hierarchy as input. }\label{table::rq4::llm}
\resizebox{\linewidth}{!}{

\begin{tabular}{lcccccccc}
\toprule
\multirow{3}{*}{\textbf{Model}} & \multicolumn{4}{c}{\textbf{Sphinx}} & \multicolumn{4}{c}{\textbf{AndroidControl}} \\
\cmidrule(lr){2-5} \cmidrule(lr){6-9}
 & \multicolumn{2}{c}{\textbf{w/o \toolname{}}} & \multicolumn{2}{c}{\textbf{with \toolname{}}} 
 & \multicolumn{2}{c}{\textbf{w/o \toolname{}}} & \multicolumn{2}{c}{\textbf{with \toolname{}}} \\
\cmidrule(lr){2-3} \cmidrule(lr){4-5} \cmidrule(lr){6-7} \cmidrule(lr){8-9}
 & \textbf{Tokens} & \textbf{SR (\%)} & \textbf{Tokens} & \textbf{SR (\%)} 
& \textbf{Tokens} & \textbf{Step SR (\%)} & \textbf{Tokens} & \textbf{Step SR (\%)} \\
\midrule
% \textbf{GPT-4o~\cite{openai2023gpt4}}  & 1,206    &   30.67   &    \textbf{596}   &   \textbf{34.36}  &   3,220    &   45.00     &    \textbf{1,484}    &    \textbf{48.00}   \\
% \textbf{Qwen-VL-Max~\cite{qwen_website}}     &   1,364  &   23.93   &    \textbf{603}    &   \textbf{31.29}     &   3,181     &   43.80    &    \textbf{1,560}    &   \textbf{45.60}    \\
% \textbf{DeepSeek-V3~\cite{liu2024deepseek}}     &   1,374    &   31.90    &   \textbf{621}     &    \textbf{36.81}    &   3,452     &    40.40    &    \textbf{1,603}    &    \textbf{42.40}    \\
% \textbf{DeepSeek-R1~\cite{guo2025deepseek}}     &  1,362   &   31.29          &  \textbf{638}   &   \textbf{34.97}   &     3,454   &    40.20    &   \textbf{1,605}  &  \textbf{45.00}  \\
% \textbf{Qwen-2.5-72B~\cite{yang2024qwen2}}    &   1,217   &   33.13  &    \textbf{594}    &   \textbf{33.74}    &   3,181     &    42.40    &    \textbf{1,560}    &   \textbf{44.40}     \\
\textbf{GPT-4o~\cite{openai2023gpt4}}  & 1,206 & 30.67 & \textbf{596 \textcolor[rgb]{0,0.5,0}{(-50.6\%)}} & \textbf{34.36} & 3,220 & 45.00 & \textbf{1,484 \textcolor[rgb]{0,0.5,0}{(-53.9\%)}} & \textbf{48.00} \\
\textbf{Qwen-VL-Max~\cite{qwen_website}} & 1,364 & 23.93 & \textbf{603 \textcolor[rgb]{0,0.5,0}{(-55.8\%)}} & \textbf{31.29} & 3,181 & 43.80 & \textbf{1,560 \textcolor[rgb]{0,0.5,0}{(-50.9\%)}} & \textbf{45.60} \\
\textbf{DeepSeek-V3~\cite{liu2024deepseek}} & 1,374 & 31.90 & \textbf{621 \textcolor[rgb]{0,0.5,0}{(-54.8\%)}} & \textbf{36.81} & 3,452 & 40.40 & \textbf{1,603 \textcolor[rgb]{0,0.5,0}{(-53.6\%)}} & \textbf{42.40} \\
\textbf{DeepSeek-R1~\cite{guo2025deepseek}} & 1,362 & 31.29 & \textbf{638 \textcolor[rgb]{0,0.5,0}{(-53.2\%)}} & \textbf{34.97} & 3,454 & 40.20 & \textbf{1,605 \textcolor[rgb]{0,0.5,0}{(-53.5\%)}} & \textbf{45.00} \\
\textbf{Qwen-2.5-72B~\cite{yang2024qwen2}} & 1,217 & 33.13 & \textbf{594 \textcolor[rgb]{0,0.5,0}{(-51.2\%)}} & \textbf{33.74} & 3,181 & 42.40 & \textbf{1,560 \textcolor[rgb]{0,0.5,0}{(-50.9\%)}} & \textbf{44.40} \\\bottomrule
\end{tabular}
}
\end{table*}

\begin{table}[t]
\centering
\caption{Comparison of different agent frameworks with and without \toolname{} on AndroidControl using Qwen-2.5-72B.}
\label{table::rq4::agent}
\begin{tabular}{lcc}
\toprule
\textbf{Agent} & \textbf{Tokens} & \textbf{Step SR (\%)} \\
\midrule
\multicolumn{3}{l}{\textit{ReAct~\cite{yao2023react}}} \\
\quad w/o UIFormer & 3,181 & 42.40 \\
\quad w/ UIFormer  & \textbf{1,560} & \textbf{44.40} \\
\midrule
\multicolumn{3}{l}{\textit{AppAgent~\cite{zhang2023appagent}}} \\
\quad w/o UIFormer & 3,457 & 41.60 \\
\quad w/ UIFormer  & \textbf{1,774} & \textbf{43.80} \\
\midrule
\multicolumn{3}{l}{\textit{Mobile-Agent-v2~\cite{wang2024mobilev2}}} \\
\quad w/o UIFormer & 3,466 & 40.40 \\
\quad w/ UIFormer  & \textbf{1,780} & \textbf{41.20} \\
\bottomrule
\end{tabular}
\end{table}

\begin{table}[t]
\caption{
Comparison of different node ordering strategies on AndroidControl. All methods use Qwen-2.5-72B and the same evaluation setup. The only difference lies in how each method formats the \toolname{}-optimized DOM tree as input.  
}
\label{table::structure}
\centering
\begin{tabular}{lcc}
\toprule
\textbf{Method} & \textbf{Hierarchy Preserved} & \textbf{Step SR}  \\
\midrule
\textbf{Random}  & \textcolor[rgb]{1,0,0}{\XSolidBrush} & 43.20 \\
\textbf{DFS}  & Flattened (Partial) & 44.00 \\
\textbf{\toolname{} (Ours)} & Fully Preserved & \textbf{44.40}  \\
\bottomrule
\end{tabular}
\end{table}

\subsubsection{Prompt analysis.}  \toolname{} preserves UI tree structure in prompts using indentation, while baseline approaches typically flatten representations. We evaluate whether this structural preservation contributes to \toolname{}'s effectiveness through an ablation study comparing different node ordering strategies on AndroidControl. Specifically, we compare two different prompt formats: \textbf{Random} ordering that removes all hierarchical structure and randomly list the nodes in the optimized UI tree, \textbf{DFS} traversal that maintains ordering but flattens representations without indentation. 
All approaches work with identical \toolname{}-optimized trees and evaluation setups, differing only in prompt formatting. 

Table~\ref{table::structure} presents the results of this ablation study. The step success rates are highly similar across all three approaches: Random (43.20\%), DFS (44.00\%), and \toolname{} (44.40\%). 
The 1.2\% absolute difference between the best and worst ablation approaches demonstrates that \toolname{}'s effectiveness mainly comes from its UI transformation programs rather than prompt formatting choices.
In addition, the results also confirm the effectiveness of \toolname{} in consolidating fragmented UI nodes into coherent semantic units, since Random ablation does not suffer from much degradation.

\subsubsection{Case study.} 
\begin{figure}
    \centering
    \includegraphics[width=0.85\linewidth]{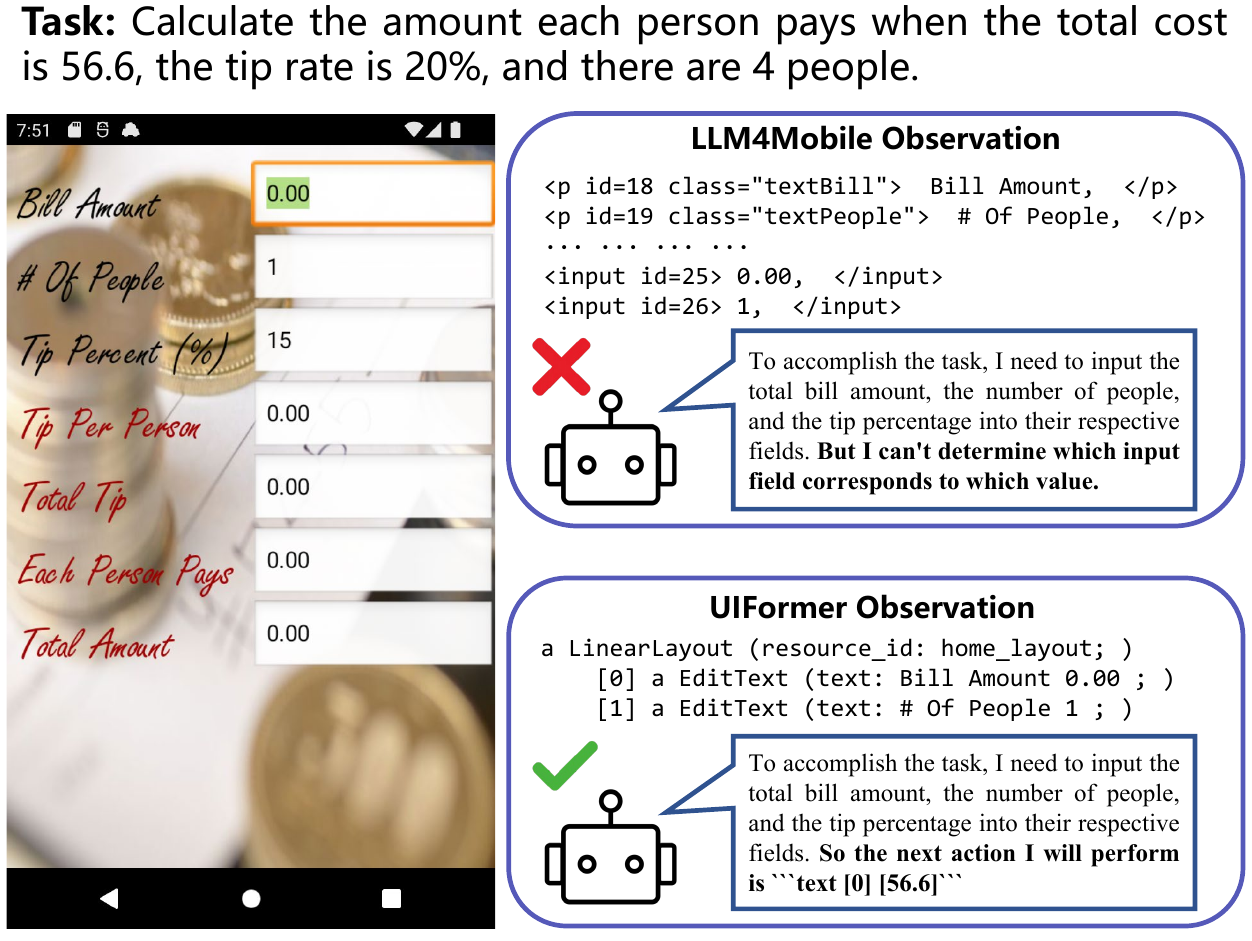}
    \caption{Case Study of Improving Agent Effectiveness.}
    \Description{A Case Study of Improving Agent Effectiveness.}
    \label{fig::eval1::case}
\end{figure}

Figure~\ref{fig::eval1::case} further demonstrates how \toolname{} can improve the agent performance  by consolidating fragmented UI nodes into coherent semantic units.
In the tip calculation task, the agent must input the total cost (56.6), tip rate (20\%), and number of people (4). The LLM4Mobile observation (i.e., Leaf representation) represents UI elements as fragmented HTML tags, where the ``Bill Amount'' label (id=18) and its corresponding input field (id=25) appear as separate, unconnected entries, preventing the LLM from determining which input field corresponds to which value and ultimately causing task failure.
\toolname{} addresses the limitation by consolidating related elements into semantic units. The bill amount label and input field are transformed into a single EditText element with descriptive content ``Bill Amount 0.00''. This semantic grouping enables the LLM to understand the functional relationship between interface components and correctly identify the target field for inputting ``56.6'' as the next action. The case study illustrates how \toolname{} transforms fragmented UI representations into coherent semantic units, reducing the cognitive burden for LLMs to reason about interface functionality.

\subsection{RQ2: Ablation on Individual Components}\label{sec::eval::ablation}
In this section, we evaluate the effectiveness of individual components in \toolname{} contributing to the overall effectiveness of \toolname{}.
Specifically, we evaluate the contribution of DSL restrictions and iterative refinement to the overall effectiveness of synthesized UI transformation programs.

\subsubsection{Direct Transformation Generation without a DSL.} To evaluate the necessity and of the DSL design, We evaluate whether a code LLM can synthesize UI transformation programs directly without a DSL to restrict the search space. Specifically, we use the same prompt shown in Figure~\ref{fig::code_synthesis_prompt} except removing the DSL specifications.
We use the program synthesized by \toolname{} as the ground-truth for checking the preservation of semantic completeness.

\begin{table}[t]
\centering
\caption{Ablation study demonstrating the necessity of DSL for program synthesis. Token inefficiency indicates additional tokens compared to optimal compression (achieved by \toolname{}); Semantic Completeness indicates whether UI semantics are preserved.}
\label{table::ablation::dsl}
\begin{tabular}{lcc}
\toprule
\textbf{Model} & \textbf{Token Inefficiency \textcolor{red}{$\downarrow$}} & \textbf{Semantic Completeness} \\
\midrule
\multicolumn{3}{l}{\textit{General-Purpose Models}} \\
\cmidrule{1-1}
GPT-4o~\cite{openai2023gpt4}         & +363\% & \textcolor[rgb]{0,0.5,0}{\Checkmark} \\
DeepSeek-V3~\cite{liu2024deepseek}   & +136\% & \textcolor[rgb]{1,0,0}{\XSolidBrush} \\
\midrule
\multicolumn{3}{l}{\textit{Reasoning Models}} \\
\cmidrule{1-1}
DeepSeek-R1~\cite{guo2025deepseek}   & +549\% & \textcolor[rgb]{0,0.5,0}{\Checkmark} \\
OpenAI o3~\cite{openaio3o4_website}                            & +203\% & \textcolor[rgb]{0,0.5,0}{\Checkmark} \\
OpenAI o4-mini~\cite{openaio3o4_website}                       & +151\% & \textcolor[rgb]{0,0.5,0}{\Checkmark} \\
Claude-Sonnet-4~\cite{claude4_website}                      & +86\%  & \textcolor[rgb]{1,0,0}{\XSolidBrush} \\
Gemini 2.5 Pro~\cite{comanici2025gemini25pushingfrontier}                       & +61\%  & \textcolor[rgb]{1,0,0}{\XSolidBrush} \\
\bottomrule
\end{tabular}
\end{table}

\noindent\textbf{Necessity of DSL Constraints.} Table~\ref{table::ablation::dsl} demonstrates that without DSL constraints, even advanced models fail to balance token efficiency and semantic completeness. Unconstrained programs exhibit two failure modes: (1) \textit{over-simplification} that fails to generalize, leading to sacrifice of semantic completeness, and (2) \textit{superficial-simplification} that fails to achieve high token efficiency. These results validate our DSL design as a necessity.

\begin{figure}[t]
    \centering
    \includegraphics[width=0.8\linewidth]{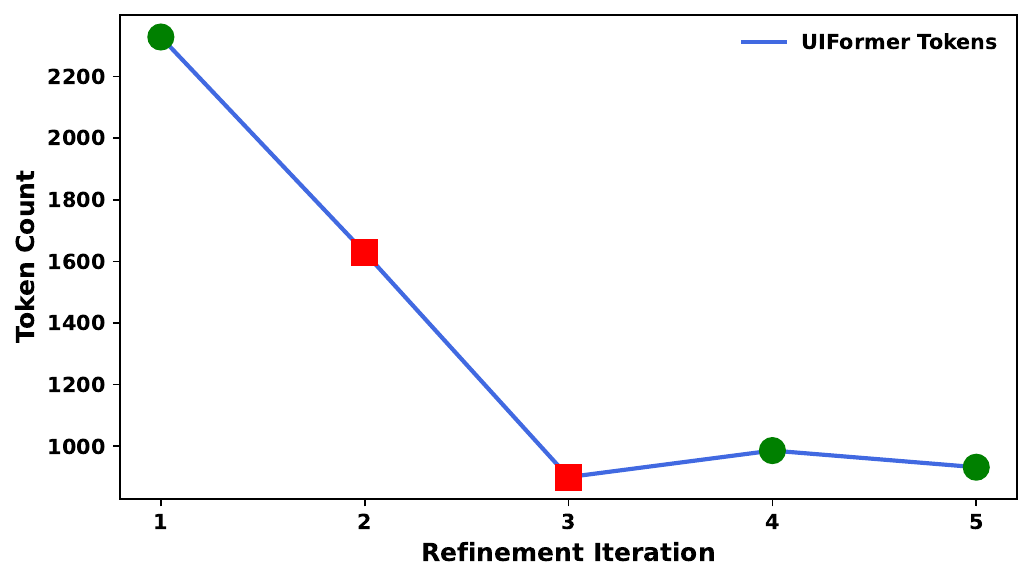}
    \caption{Token reduction of achieved by the programs synthesized by \toolname{} in different iterations. Red squares indicate generated programs violating semantic completeness.}
    \Description{A line or bar chart showing token reduction across multiple synthesis iterations of \toolname{}. Most points indicate progressive improvement, while red square markers highlight cases where the generated programs failed to preserve semantic completeness.}
    \label{fig:iterative}
\end{figure}

\subsubsection{Ablation on Iterative Refinement.}
we conduct an ablation study to study the impact of iterative refinement. Specifically, we use the same experimental setup and DSL constraints as \toolname{}, but evaluate the initial program generated by each model without any subsequent modifications.

\noindent\textbf{Iterative Refinement Benefits.} Figure~\ref{fig:iterative} demonstrates that iterative refinement successfully balances token efficiency and semantic correctness. The initial program generates 2,280 tokens while preserving semantic completeness. 
Early refinement iterations (2-3) achieve aggressive token reduction but violate semantic completeness (red triangles), producing programs that lose critical UI information. However, receiving the feedback, subsequent iterations (4-5) recover semantic correctness while maintaining significant compression, achieving aobut 60\% token reduction compared to the initial program.

\subsection{RQ3: Practicality for Industry Deployment}

To evaluate the real-world effectiveness and practicality of \toolname{}, we deploy \toolname{} at the Development and Engineering Tools (DET) team of WeChat, an international IT company maintaining highly popular industrial apps with over \textbf{one billion} monthly active users.
The DET team provides LLM-based testing services for in-house testing of Android and Web applications.
The company operates LLM agents that provide automated GUI testing services, where each request comprises a natural language instruction paired with a UI tree representation, and the system generates appropriate UI actions in response.
We integrate \toolname{} as a preprocessing component in the existing serving pipeline, where UI trees are transformed by \toolname{} before being fed to the LLM. 
The service typically handles workloads with peaks of up to 200 concurrent requests using 32 H20 GPUs.

\textbf{Evaluation setting.} To quantify efficiency improvements, we record all requests processed during a one-hour period and subsequently replay this workload under two configurations: (1) baseline serving without \toolname{} preprocessing, and (2) enhanced serving with \toolname{} UI transformation. 
Both configurations utilized identical hardware infrastructure consisting of 8 NVIDIA H20 GPUs (768GB memory in total) hosting a Qwen-2.5 72B model through the vLLM serving framework~\cite{kwon2023efficient}.
We measure the quality of service (QoS) of the LLM agent inference service with latency (the time interval between requesting the service and receiving the response), and throughput in queries per minute (QPM), i.e., the number of queries that the service can complete.

\begin{table}[t]
\centering
\caption{Quality of Service (QoS) comparison of LLM agent serving with and without \toolname{} under 50 concurrent requests. \textcolor{red}{$\downarrow$} and \textcolor[rgb]{0,0.5,0}{$\uparrow$} indicate lower and higher is better, respectively.}
\label{table::eval::industry::serving}
\begin{tabular}{lrrr}
\toprule
 \multicolumn{1}{l}{\textbf{QoS Metric}} & \multicolumn{1}{c}{\textbf{Baseline}} & \multicolumn{1}{c}{\textbf{+ \toolname{}}} & \multicolumn{1}{c}{\textbf{$\Delta$}} \\ \hline
\textbf{Avg. \# of tokens \textcolor{red}{$\downarrow$}} & 8,685 & 2,010 & \textbf{-76.9\%} \\
\textbf{Min. latency (ms) \textcolor{red}{$\downarrow$}} & 3,316 & 402 & \textbf{-87.9\%} \\
\textbf{Max. latency (ms) \textcolor{red}{$\downarrow$}} & 23,125 & 9,717 &\textbf{-58.0\%} \\
\textbf{Avg. latency (ms) \textcolor{red}{$\downarrow$}} & 7,131 & 5,273 &\textbf{ -26.1\%} \\
\textbf{Throughput (QPM) \textcolor[rgb]{0,0.5,0}{$\uparrow$}} & 421 & 569 &\textbf{ +35.2\%} \\ 
\bottomrule
\end{tabular}
\end{table}

\begin{figure}[t]
  \centering
  \includegraphics[width=0.85\linewidth]{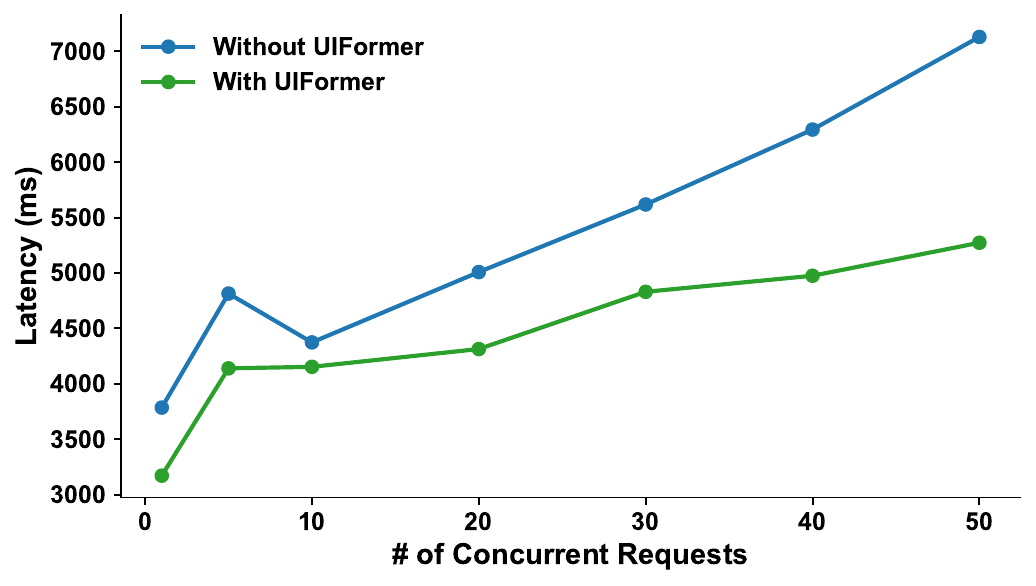}
  \caption{Average latency of the LLM agent service with and without \toolname{} under different levels of concurrency.}
  \Description{A plot comparing the average response latency of the LLM agent service. 
The x-axis shows different concurrency levels, while the y-axis shows average latency in milliseconds or seconds. 
Two lines are plotted: one representing the baseline service without \toolname{}, and another with \toolname{}. 
The curve with \toolname{} generally maintains lower latency as concurrency increases.}
  \label{fig::latency_scalability}
\end{figure}

\textbf{Main results.} Table~\ref{table::eval::industry::serving} presents the QoS metrics obtained from our deployment evaluation. 
The integration of \toolname{} yields substantial improvements across all measured dimensions of QoS and efficiency. 
\toolname{} achieves a 76.9\% reduction in token consumption per request on average, directly contributing to latency reduction by 26.1\% on average and a 35.2\% increase in service throughput. 
These results demonstrate \toolname{}'s effectiveness in improving service quality for real-world services, confirming the practicality and values in industry deployment. In addition, \toolname{} makes the service more scalable.
As shown in Figure~\ref{fig::latency_scalability}, with the increase of concurrent requests, \toolname{} maintains consistently lower latency and exhibits better scalability with the increase of concurrency requests. 
While the baseline service experiences steep latency degradation under increasing load, \toolname{}-enhanced service demonstrates more graceful performance scaling, maintaining reasonable response times even at high concurrency levels.

\section{Related Work}\label{sec::related}

\subsection{LLM Agents for UI Navigation}
Existing LLM agents for UI navigation primarily fall into two categories.
(1) Prompt-based approaches combine domain-specific UI knowledge with LLMs through specialized mechanisms.
Wang et al.\ \cite{wang2023enabling} represent UI as HTML and using few-shot prompts to map instructions to UI actions.
AutoDroid~\cite{wen2024autodroid} enhances performance by leveraging UI transition graphs and embedding models to retrieve similar task examples as in-context demonstrations.
Guardian~\cite{ran2024guardian} proposes offloading computational demands from LLMs to specialized modules with domain knowledge, improving planning capabilities.
Other works~\cite{zhang2023appagent, wang2024mobile, wang2024mobilev2, yan2023gpt} explore navigation agents that utilize UI screenshots combined with set-of-mark prompt techniques.
(2) Training-based approaches develop UI-specific models.
Grounding models\cite{yang2024aria, gou2024navigating} are trained to work alongside proprietary planning models that generate low-level instructions, helping locate UI elements requiring interaction.
UI navigation models\cite{hong2024cogagent, cheng2024seeclick, ma2024coco, lin2024showui, wu2024atlas, qin2025ui} leverage specialized datasets to enhance planning capabilities. While existing work heavily focuses on improving agent reasoning and planning capabilities, \toolname{} addresses an orthogonal problem, i.e., UI representation efficiency, improving existing work for efficient use.

\subsection{UI Representations for UI Agents}
Existing researches use screenshot~\cite{wan2024omniparser, song2024visiontasker, zhang2023appagent, wang2024mobile, wang2024mobilev2, yan2023gpt, rawles2024androidworld, taeb2024axnav} or text representations derived from UI tree ~\cite{wang2023enabling,droidGPT,wen2024autodroid,ran2024guardian,vu2024gptvoicetasker} to represent UIs to LLM agents.
However, as noted in OSWorld~\cite{xie2024osworld} and WebArena~\cite{zhou2023webarena}, screenshot-only representations currently demonstrate lower performance, and we focus on text representations in this work.
Given the complexity of UI trees, existing approaches adopts different heuristics to simplify them.
LLM4Mobile~\cite{wang2023enabling} preserves all leaf nodes and categorizes them into five classes to generate HTML representations that better align with training data distributions.
Similarly, AutoDroid~\cite{wen2024autodroid} employs HTML formatting for leaf nodes while additionally pruning equivalent elements based on UI transition graphs and hierarchy structures.
Guardian~\cite{ran2024guardian} extracts only interactive elements from UI hierarchies to generate concise lists, whereas GPTVoiceTasker~\cite{vu2024gptvoicetasker} utilizes tree structures for UI representation.
Instead of proposing heuristics for UI representation simplification, \toolname{} is an automated and principled approach to co-optimize token efficiency and semantic completeness.

\subsection{Efficiency-oriented Optimization for LLMs}
The efficiency-oriented optimization for LLMs emerges as an important problem, with the primary focus on LLM-based code generation~\cite{suneja2021probingmodelsignalawarenesspredictionpreserving, Rabin_2022syntax,  chaoxuan2023structual,Saad_2025alpine}.
By extracting key partial information as inputs for LLMs, existing approaches~\cite{suneja2021probingmodelsignalawarenesspredictionpreserving, Rabin_2022syntax, chaoxuan2023structual,Saad_2025alpine} can considerably boost efficiency without significantly compromising performance of code models.
Sivand~\cite{Rabin_2021sivand} first demonstrates that it can remove more than 60\% of the input program while preserving the predictions of code model.
DietCode~\cite{Zhang_2022dietcode} builds an attention dictionary for code tokens and code statements, pruning the insignificant ones and reducing the input length to 60\%, with a slight performance drop.
SIMPY~\cite{sun2024aicodersusrethinking} introduces a simplified grammar for Python, achieving a reduction in token usage by over 10\% without changing the program's abstract syntax tree. 
While existing work demonstrate the necessity of code LLM efficiency, \toolname{} tackles a different problem for UI agent efficiency and we designs novel solution, i.e., DSL-based iterative refinement to effectively hardness the code LLM for high-quality transformation program synthesis.
\section{Threats to Validity}\label{sec::threats}

The threats to internal validity concern instrumentation effects, such as implementation faults in \toolname{} and potential errors in scripts used to generate experimental results. Additionally, incorrect configuration of baselines could have affected our findings. To mitigate the threats, we cross-validate our implementation and use publicly available implementations of baselines with their default configurations, maintain detailed experimental logs, and manually analyze representative samples to ensure correct setup. The external threat concerns the representativeness of our experimental subjects, including the selected Android applications, UI platforms, LLMs, and LLM agents. Our findings may not generalize to applications, platforms, or models beyond those evaluated in our study. We address this limitation by evaluating across diverse application domains and multiple state-of-the-art language models and agents.
The threat to construct validity concerns randomness inherent in both Android application behavior and LLM inference processes. To mitigate this threat, we employ aggregation metrics across multiple experimental runs and use low temperature settings for LLM inference to reduce variability in model outputs.
\section{Conclusion}
LLM-based UI agents have been increasingly adopted for complex UI navigation tasks, but their efficiency remains an underexplored challenge. Despite the promising capabilities of LLMs in understanding UI semantics, we have found two major inefficiencies in existing approaches: the systematic fragmentation patterns in UI implementations that create unnecessarily complex representation spaces, and the lack of automated optimization mechanism to improve the token efficiency without sacrificing semantic completeness.
To address the preceding challenges, in this paper, we have proposed \toolname{}, an automated optimization framework with two key designs. First, we design a DSL to constrain the search space.
Second, \toolname{} incorporates LLM-based iterative refinement with local feedback on semantic completeness and token efficiency.
We have constructed comprehensive evaluations across multiple UI domains and deployed \toolname{} in industrial settings. Evaluation results show that \toolname{} achieves 48.7\% to 55.8\% reduction in token usage while maintaining semantic completeness. Further experiments and real-world deployment have confirmed the effectiveness and practicality of \toolname{}.

\bibliographystyle{ACM-Reference-Format}
\bibliography{ref}
\balance

\end{document}